\documentclass[letterpaper, 10 pt, conference]{ieeeconf}  

\IEEEoverridecommandlockouts                              

\usepackage{graphics} 

\usepackage{amsmath,amssymb,amsfonts}
\usepackage{mathtools}
\newcommand{\B}[1]{#1} 

\newtheorem{thm}{Theorem}[section]
\newtheorem{lem}[thm]{Lemma}

\newtheorem{rem}{Remark}

\usepackage[style=ieee]{biblatex}
\usepackage{algorithm, algpseudocode}
\algnewcommand{\LineComment}[1]{\Statex \(\triangleright\) #1 \(\triangleleft\)}

\usepackage{hyperref}

\usepackage{xcolor}

\title{\LARGE \bf
Stability Margins of Neural Network Controllers*
}

\author{Neelay Junnarkar$^{1}$, Murat Arcak$^{1}$, and Peter Seiler$^{2}$
\thanks{*Supported by NFS CNS-2111688 and AFOSR FA9550-23-1-0732.}
\thanks{$^{1}$Neelay Junnarkar and Murat Arcak are with the Department of Electrical Engineering and Computer Sciences, University of California, Berkeley, CA 94720 USA ({\tt neelay.junnarkar@berkeley.edu, arcak@berkeley.edu})}%
\thanks{$^{2}$Peter Seiler is with the Department of Electrical Engineering and Computer Science, University of Michigan, Ann Arbor, MI 48109 USA ({\tt pseiler@umich.edu})}%
}

\begin{document}

\maketitle
\thispagestyle{empty}
\pagestyle{empty}

\begin{abstract}

We present a method to train neural network controllers with guaranteed stability margins.
The method is applicable to linear time-invariant plants interconnected with uncertainties and nonlinearities that are described by integral quadratic constraints.
The type of stability margin we consider is the disk margin.
Our training method alternates between a training step to maximize reward and a stability margin-enforcing step.
In the stability margin enforcing-step, we solve a semidefinite program to project the controller into the set of controllers for which we can certify  the desired disk margin.

\end{abstract}
\section{Introduction}

Neural network controllers trained through standard learning techniques show promise to overcome limitations of traditional  control designs, but provide no guarantees on closed-loop behavior.
This has limited the use of neural networks in safety-critical applications, such as aerospace, where it is common for regulatory agencies to formulate controller robustness requirements in terms of stability margins, such as classical gain and phase margins \cite{faa}.

Recent work on safe learning has focused on verification and synthesis of neural network controllers.
References \cite{yinStabilityAnalysisUsing2022, hashemiCertifyingIncrementalQuadratic2021, pauliLinearSystemsNeural2021} 
verify asymptotic stability and 
find inner approximations to 
regions of attraction of feedback systems involving neural networks. 
Methods that synthesize neural network controllers and jointly certify closed-loop stability include \cite{Gu_Yin_Ghaoui_Arcak_Seiler_Jin_2022, wangLearningAllStabilizing2023, junnarkarSynthesisStabilizingRecurrent2022, junnarkar2024synthesizingneuralnetworkcontrollers}.
These methods differ in the types of plants considered, with \cite{wangLearningAllStabilizing2023} considering linear time-invariant (LTI) plants, \cite{junnarkarSynthesisStabilizingRecurrent2022} 
addressing
LTI plants with sector-bounded uncertainties, and \cite{Gu_Yin_Ghaoui_Arcak_Seiler_Jin_2022, junnarkar2024synthesizingneuralnetworkcontrollers} 
studying LTI plants interconnected with uncertain blocks described by integral quadratic constraints.

Besides being used as controllers, neural networks have also been explored as system models.
Reference \cite{barabanovStabilityAnalysisDiscretetime2002} uses linear matrix inequality techniques to analyze stability of an equilibrium of a recurrent neural network.
Reference \cite{karnyRecurrentNeuralNetworks1998} analyzes approximation capabilities, controllability, and observability of recurrent neural networks.
These results can be 
used for
verifying closed-loop properties of a neural network controller 
by modeling the interconnection of the controller and plant as a neural network.
However, they are less
amenable to neural network controller synthesis due to the structure 
imposed
by the fixed plant parameters.

In safety-critical applications, such as aviation, the classical notions of 
gain and phase margins are well-accepted to the point of being a part of regulations of the Federal Aviation Administration \cite{faa}.
Satisfaction of a 
stability margin ensures the controller will stabilize a plant despite 
a class of unmodeled dynamics.
However, it is possible for systems with large gain and phase margins to be destabilized by an arbitrarily small simultaneous variation in plant gain and phase \cite{zhouRobustOptimalControl1996}.
Instead, the notion of {\it disk margin} \cite{seilerIntroductionDiskMargins2020} 
allows
simultaneous gain and phase variations.
Although the disk margin accounts for LTI uncertainty 
described 
in the
frequency domain, 
it can be 
reinterpreted in the time domain and generalized to nonlinear and time-varying plant and controller models.
This has been used in \cite{schugRobustnessMarginsLinear2017} to assess disk margins of linear parameter-varying feedback systems.

In this paper, we present a method to train a neural network controller to maximize a reward subject to guaranteed disk margins.
We leverage integral quadratic constraints to describe the disk margin and the neural network activation functions in time domain.
We derive linear matrix inequality conditions for Lyapunov stability 
and translate them to
semidefinite programs that enforce stability margins after each training step in the reinforcement learning process.
A neural network trained through this method can be used in place of a traditional controller 
when
stability margins are 
required.
This allows improving an auxiliary performance metric, used as the reward in training, while satisfying the same robustness constraints as the traditional controller.


In Section~\ref{sec:problem-setup}, we provide background on disk margins and our neural network model, and provide the problem statement.
In Section~\ref{sec:certification}, we 
characterize
the disk margin of the closed-loop system with a neural network controller.
In Section~\ref{sec:training}, we present a method to train a neural network controller to maximize reward subject to the closed-loop satisfying a disk margin.
In Section~\ref{sec:experiments}, we demonstrate our method in simulation for a flexible rod on a cart.

\subsection{Notation}
$\mathbb{R}_{\geq 0}$ denotes the set of nonnegative real numbers, $\mathcal{L}_2^n$ the space of square-integrable functions from $\mathbb{R}_{\geq 0}$ to $\mathbb{R}^n$, and  $\mathcal{L}_{2e}^n$ the space of functions from  $\mathbb{R}_{\geq 0}$ to $\mathbb{R}^n$ that are square-integrable on 
$[0, T]$ for all $T > 0$.
We drop the superscript 
for
the dimension of the codomain 
when
clear from context.
We use $(\star)^\top X y$ to denote $y^\top X y$, and $\mathrm{He}(y)$ to denote $y + y^\top$.

\section{Background and Problem Setup} \label{sec:problem-setup}

We consider the problem of robustly stabilizing a plant using a neural network controller, with 
the further
objective of maximizing a reward.
We first describe the plant uncertainty and  neural network, and then give the problem formulation.

\subsection{Plant Model and Disk Margins}


Consider the nominal plant, $P$:
\begin{equation} \label{eq:plant} 
\begin{aligned}
    \begin{bmatrix} \dot{x}_{p}(t) \\ y(t) \end{bmatrix}
    & =
    \begin{bmatrix}
        A_p  & B_p \\
        C_p & 0
    \end{bmatrix}
    \begin{bmatrix} x_p(t) \\ u(t) \end{bmatrix}
\end{aligned}
\end{equation}
where $x_p(t) \in \mathbb{R}^{n_p}$, $y(t) \in \mathbb{R}^{n_y}$, $u(t) \in \mathbb{R}^{n_u}$ are the plant state and output, and control input at time $t$. 
 Although we study a LTI plant for brevity, the results extend
to plants with nonlinearities described by integral quadratic constraints.


\begin{figure}[tb]
    \centering
    \includegraphics[scale=0.2]{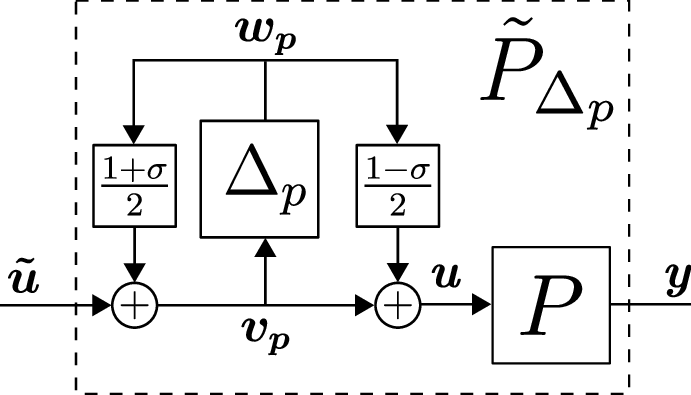}
    \caption{
        Block-diagram of disk margin uncertainty at plant input.
    }
    \label{fig:disk-margin}
\end{figure}

We describe robustness to uncertainty with a disk margin.
The traditional disk margin $D(\alpha,\sigma)$ accounts for LTI input uncertainty whose frequency response, in each channel, lies in the disk parameterized by $\alpha \in \mathbb{R}_{\geq 0}$ and $\sigma \in \mathbb{R}$:
\begin{align*} 
  \left\{ 
  \frac{1+ \frac{1-\sigma}{2} \delta }{1- \frac{1+\sigma}{2}
  \delta} :
  \delta \in \mathbb{C}  \mbox{ with } |\delta| < \alpha \right\}.
\end{align*}
The value $\alpha$ determines the size of the uncertainty disk.
Note that the disk margin can be used to compute lower bounds on the classical gain and phase margins.
For example, if the system remains stable for $D(\alpha, \sigma)$, then the system is stable for all gain variations in the range $[\gamma_{min},\gamma_{max}]$ where
\begin{align*}
\gamma_{\min} = \frac{2-\alpha(1-\sigma)}{2+\alpha(1+\sigma)} \mbox{ and }
\gamma_{\max} = \frac{2+\alpha(1-\sigma)}{2-\alpha(1+\sigma)}.
\end{align*}

The disk margin definition above can be interpreted as robustness to input unmodeled dynamics described by the class of transfer functions 
$$H_{\Delta_p}(s) = \left(I + \frac{1-\sigma}{2} \Delta_p(s)\right)\left(I + \frac{1+\sigma}{2} \Delta_p(s)\right)^{-1}$$ where $\Delta_p$ is LTI, diagonal,
and $\|\Delta_p\|_{\mathcal{H}_\infty} < \alpha$.
The choice $\Delta_p(s)=0$ gives $H_{\Delta_p}(s)=I$, 
which corresponds to the nominal system.
The input uncertainty $H_{\Delta_p}(s)$ is represented in block-diagram form in Figure~\ref{fig:disk-margin}, 
where \begin{equation} \label{eq:Ptilde_Delta}
\begin{alignedat}{2}
    y & = P u, & \quad
    u & = v_p + \frac{1-\sigma}{2} w_p, \\
    v_p & = \tilde{u} + \frac{1+\sigma}{2} w_p, & \quad
    w_p & = \Delta_p v_p.
\end{alignedat}
\end{equation}
The uncertain system from input $\tilde{u}$ to output $y$, for a particular $\Delta_p$, is denoted by $\tilde{P}_{\Delta_p}$.

This interpretation allows us to broaden the disk margin to non-LTI uncertainty by generalizing $\Delta_p$ to be $\mathcal{L}_2$ norm-bounded by $\alpha$.
The class of plants which must be stabilized is then $\mathcal{M}_\alpha = 
\{\tilde{P}_{\Delta_p} \; | \; \mathcal{L}_2\ \text{gain of}\ \Delta_p\ \text{is} < \alpha \}$.
In the rest of this paper, we say that a controller satisfies the disk margin $D(\alpha, \sigma)$ if it stabilizes the class of plants $\mathcal{M}_\alpha$.

\subsection{Neural Network Controller Model}

\begin{figure}[tb]
    \centering
    \includegraphics[scale=0.2]{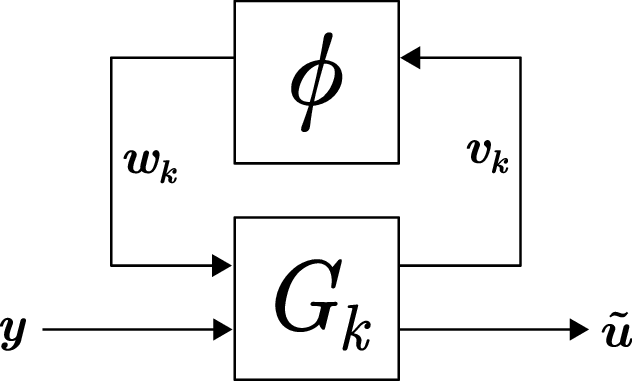}
    \caption{
        The neural network controller is modeled as the interconnection of an LTI system $G_k$ and activation functions $\phi$.
    }
    \label{fig:K}
\end{figure}

We model the neural network controller as the interconnection of an LTI system and activation functions (Figure~\ref{fig:K}):
\begin{equation} \label{eq:controller} 
\begin{aligned}
    \begin{bmatrix} \dot{\B{x}}_{\B{k}}(t) \\ \B{v_k}(t) \\ \B{\tilde{u}}(t) \end{bmatrix}
    & = 
    \begin{bmatrix}
        A_k & B_{k w} & B_{k y} \\
        C_{k v} & D_{k v w} & D_{k v y} \\
        C_{k u} & D_{k u w} & D_{k u y}
    \end{bmatrix}
    \begin{bmatrix}
        \B{x_k}(t) \\ \B{w_k}(t) \\ \B{y}(t)
    \end{bmatrix}, \\
    \B{w_k}(t) & = \phi(\B{v_k}(t)),
\end{aligned}
\end{equation}
where $x_k(t) \in \mathbb{R}^{n_k}$, $v_k(t) \in \mathbb{R}^{n_\phi}$, and $w_k(t) \in \mathbb{R}^{n_\phi}$ are, respectively, the controller state, the input to the activation function $\phi$, and the output of the activation function $\phi$ at time $t$.
Many neural networks can be modeled in this form, though we restrict ourselves to neural networks without biases to ensure that the controller does not shift the equilibrium.

The activation function $\phi$ is memoryless and applied elementwise:
$w_{k, i} = \phi_i(v_{k, i})$.
Each scalar activation function $\phi_i$ is sector-bounded in $[0, 1]$, and slope-restricted in $[0, 1]$.
Common activation functions that satisfy the sector bounds and slope restrictions are tanh and ReLU.

The controller \eqref{eq:controller} is well-posed if there exists a unique solution for $w_k$ in the implicit equation $w_k = \phi(C_{k v} x_k + D_{k v w} w_k + D_{k v y} y_k)$ for all $x_k$ and $y_k$.
The system is trivially well-posed if $D_{k v w} = 0$ but otherwise requires constraints on $D_{k v w}$.
We refer to \cite{junnarkar2024synthesizingneuralnetworkcontrollers} for a discussion of this issue.

When $n_k = 0$, this controller model simplifies to $\B{\tilde{u}}(t) = D_{k u w} \B{w_k}(t) + D_{k u y} \B{y}(t)$ where $\B{w_k}(t) = \phi(D_{k v w}\B{w_k}(t) + D_{k v y} \B{y}(t))$.
This is a static implicit neural network (INN), which encompasses common classes of neural networks, including fully connected feedforward networks, convolutional layers, max-pooling layers, and residual networks \cite{elghaouiImplicitDeepLearning2021}.

For example, consider the feedforward neural network:
\begin{align*}
    \tilde{u} = W_L w_L + b_L, w_{l+1} = \phi_l(W_l w_l + b_l), w_0 = y.
\end{align*}
With $w_k = \begin{bmatrix} w_1 & \cdots & w_L \end{bmatrix}^\top$ and $\phi = \begin{bmatrix} \phi_0 & \cdots & \phi_{L-1}\end{bmatrix}^\top$, this can be represented by the following INN:
\begin{align*}
    \tilde{u} & = \underbrace{\begin{bmatrix} 0 & \cdots & 0 & W_l \end{bmatrix}}_{D_{kuw}} w_k + \underbrace{\begin{bmatrix} 0 & b_L \end{bmatrix}}_{D_{kuy}} \begin{bmatrix} y \\ 1 \end{bmatrix} \\
    w & = \phi\left(
    \underbrace{\begin{bsmallmatrix}
        0 \\
        W_1 & 0  \\
        0 & W_2 & 0 \\
        \vdots & \ddots & \ddots & \ddots & \\
        0 & \cdots & 0 & W_{L-1} & 0
    \end{bsmallmatrix}}_{D_{kvw}} w_k
    + \underbrace{\begin{bsmallmatrix}
        W_0 & b_0 \\
        0 & b_1 \\
        \vdots & \vdots \\
        0 & b_{L-1}
    \end{bsmallmatrix}}_{D_{kvy}} \begin{bmatrix} y \\ 1 \end{bmatrix}
    \right).
\end{align*}

When $n_k > 0$, the controller has dynamics, making
the controller \eqref{eq:controller} a \textit{recurrent} implicit neural network (RINN).
Controllers with memory are useful in applications with partially-observed plants \cite{callierLinearSystemTheory1991}.
This model and variations have been used for control in \cite{junnarkar2024synthesizingneuralnetworkcontrollers, Gu_Yin_Ghaoui_Arcak_Seiler_Jin_2022, junnarkarSynthesisStabilizingRecurrent2022, wangLearningAllStabilizing2023}.

\subsection{Problem Statement}

We now address the constrained optimization problem:
\begin{equation}
\begin{aligned}
    \max_K & \quad \mathbb{E}\left[\int_{0}^T r(\B{x_p}(t), \B{u}(t)) dt\right] \\
    \text{s.t.} & \quad \mathcal{F}(\tilde{P}_{\Delta_p}, K)\ \text{stable}\ \forall \tilde{P}_{\Delta_p} \in \mathcal{M}_\alpha,\\
    & \quad K\ \text{is of the form in \eqref{eq:controller}}
\end{aligned}
\end{equation}
where $\mathcal{F}(\tilde{P}_{\Delta_p}, K)$ denotes the closed-loop of $\tilde{P}_{\Delta_p}$ and $K$, and $r: \mathbb{R}^{n_p} \times \mathbb{R}^{n_u} \to \mathbb{R}$ is the reward function.
The integral of the reward over a trajectory on $[0, T]$, where $T \in \mathbb{R}_{\geq 0}$, gives a score to the trajectory.
The expectation is taken over a distribution of initial plant states, and any randomness in the training model (for example, disturbances).
Note that the simulation model and environment used for rolling out trajectories in training need not be the same as the nominal plant model; it can be a higher-fidelity simulation model.

\section{Disk Margin Certification} \label{sec:certification}

In this section, we present a condition certifying that a RINN controller $K$ satisfies the disk margin $D(\alpha, \sigma)$ for a nominal plant $P$.


\subsection{Transforming Plant Model}

We transform the uncertain plant $\tilde{P}_{\Delta_p}$ into the form of an LTI system interconnected with an uncertainty block.
This form mirrors the form of the RINN controller, and enables the entire closed-loop system to be modeled as an LTI system interconnected with an uncertainty block (grouping the controller's  activation functions into the uncertainty block).

We simplify the LTI part of the uncertain plant \eqref{eq:Ptilde_Delta}, making an LTI system with inputs $\B{w_p}$ and $\B{\tilde{u}}$ and outputs $\B{v_p}$ and $\B{y}$.
Combining $u = v_p + \frac{1 - \sigma}{2} w_p$ and $v_p = \tilde{u} + \frac{1 + \sigma}{2} w_p$ gives $\B{u} = \B{w_p} + \B{\tilde{u}}$.
Plugging this into \eqref{eq:plant}:
\begin{align*}
    \begin{bmatrix} \dot{\B{x}}_{\B{p}}(t) \\ \B{y}(t) \end{bmatrix}
    & =
    \begin{bmatrix}
        A_p  & B_p \\
        C_p & 0
    \end{bmatrix}
    \begin{bmatrix} \B{x_p}(t) \\ \B{u}(t) \end{bmatrix} \\
    & = \begin{bmatrix}
        A_p & B_p & B_p \\ C_p & 0 & 0
    \end{bmatrix}
    \begin{bmatrix} \B{x_p}(t) \\ \B{w_p}(t) \\ \B{\tilde{u}}(t) \end{bmatrix}.
\end{align*}

We next add $\B{v_p}$ as an output using $\B{v_p}(t)$ from \eqref{eq:Ptilde_Delta}:
\begin{equation} \label{eq:lft-plant}
    \begin{aligned}
        \begin{bmatrix} \dot{\B{x}}_{\B{p}}(t) \\ \B{v_p}(t) \\ \B{y}(t) \end{bmatrix}
        & = 
        \begin{bmatrix}
        A_p & B_p & B_p \\ 0 & \frac{1+\sigma}{2} I & I \\ C_p & 0 & 0
        \end{bmatrix}
        \begin{bmatrix} \B{x_p}(t) \\ \B{w_p}(t) \\ \B{\tilde{u}}(t) \end{bmatrix} \\
        \B{w_p}(t) & = {\Delta_p}(\B{v_p})(t).
    \end{aligned}
\end{equation}
Equation \eqref{eq:lft-plant} gives the plant $\tilde{P}_{\Delta_p}$ in the form of an LTI system interconnected with an uncertainty, similar to the controller diagram in Figure~\ref{fig:K}, as opposed to the structured interconnection in Figure~\ref{fig:disk-margin}.

\subsection{Quadratic Constraints (QCs) on $\Delta_p$ and $\phi$}

With both the plant and the controller now in the form of an LTI system interconnected with an uncertainty or a nonlinearity, we characterize the non-LTI components---that is, ${\Delta_p}$ and $\phi$---by describing the relationship between the inputs and outputs of each block with quadratic constraints \cite{megretskiSystemAnalysisIntegral1997, seilerStabilityAnalysisDissipation2015, schererDissipativityIntegralQuadratic2022, veenmanIQCsynthesisGeneralDynamic2014}.
For simplicity and computational tractability, we use static quadratic constraints for both ${\Delta_p}$ and $\phi$.

Since ${\Delta_p}$ is diagonal and has $\mathcal{L}_2$ gain less than or equal to $\alpha$, it satisfies the following integral quadratic constraint (IQC) for all $\Lambda_p \in \mathbb{R}^{n_u \times n_u}$ where $\Lambda_p \succeq 0$ and diagonal, $\B{v_p} \in \mathcal{L}_{2e}^{n_u}$, $\B{w_p} = {\Delta_p}(\B{v_p})$, and $T \geq 0$ \cite{megretskiSystemAnalysisIntegral1997}:
\begin{equation} \label{eq:deltap-iqc}
    \int_0^T \left.
    \begin{bmatrix} \B{v_p}(t) \\ \B{w_p}(t) \end{bmatrix}^\top
    \begin{bmatrix} \alpha^2 \Lambda_p & 0 \\ 0 & -\Lambda_p \end{bmatrix}
    \begin{bmatrix} \B{v_p}(t) \\ \B{w_p}(t) \end{bmatrix}
    \right.
    \geq 0.
\end{equation}
To restrict $\Delta_p$ to be LTI, one can include dynamic filters into \eqref{eq:deltap-iqc}.
The class of filters for uncertain LTI dynamics is described in \cite{megretskiSystemAnalysisIntegral1997}.

Since $\phi$ is applied elementwise and sector-bounded in $[0, 1]$, it satisfies the following quadratic constraint for all $\Lambda_k \in \mathbb{R}^{n_\phi \times n_\phi}$ where $\Lambda_k \succeq 0$ and diagonal, $v_k \in \mathbb{R}^{n_\phi}$, and $w_k = \phi(v_k)$ \cite{fazlyabSafetyVerificationRobustness2022}:
\begin{equation}
    \begin{bmatrix} v_k \\ w_k \end{bmatrix}^\top
    \begin{bmatrix} 0 & \Lambda_k \\ \Lambda_k & -2 \Lambda_k \end{bmatrix}
    \begin{bmatrix} v_k \\ w_k \end{bmatrix}
    \geq 0.
\end{equation}

\subsection{Stability Condition}

We form the closed-loop system of \eqref{eq:lft-plant} and \eqref{eq:controller}:
\begin{equation} \label{eq:closed-loop} 
    \begin{aligned}
        \begin{bmatrix} \dot{\B{x}}(t) \\ \B{v}(t) \end{bmatrix} 
        & = \begin{bmatrix}
            A & B_w  \\
            C_v & D_{v w} \\
        \end{bmatrix}
        \begin{bmatrix} \B{x}(t) \\ \B{w}(t) \end{bmatrix} \\
        \B{w}(t) & = \Delta(\B{v})(t),
    \end{aligned}
\end{equation}
where $x(t) = \begin{bmatrix} x_p(t)^\top & x_k(t)^\top \end{bmatrix}^\top$, $v(t) = \begin{bmatrix} v_p(t)^\top & v_k(t)^\top \end{bmatrix}^\top$, $w(t) = \begin{bmatrix} w_p(t)^\top & w_k(t)^\top \end{bmatrix}^\top$, and $\Delta: \B{v} \mapsto \begin{bmatrix} \Delta_p(\B{v_p})^\top & \phi(\B{v_k})^\top \end{bmatrix}^\top$.
The expansions of $A, B_w, C_v, D_{v w}$ matrices are:
\begin{align*}
   A & = \begin{bsmallmatrix}
        A_p + B_{p} D_{k u y} C_{p} & B_{p} C_{k u} \\
        B_{k y} C_{p} & A_k
    \end{bsmallmatrix} &
    B_w  = \begin{bmatrix}
        B_{p} & B_{p} D_{k u w} \\
        0 & B_{k w}
    \end{bmatrix} \\
    C_v & = \begin{bmatrix}
        D_{k u y} C_{p} & C_{k u} \\
        D_{k v y} C_{p} & C_{k v}
    \end{bmatrix} &
    D_{v w}  = \begin{bmatrix}
        \frac{1+\sigma}{2} I & D_{k u w} \\
        0 & D_{k v w}
    \end{bmatrix}. \\
\end{align*}

Since $\Delta$ is formed by stacking $\Delta_p$ and $\phi$, it satisfies an integral quadratic constraint formed by interleaving the IQCs satisfied by $\Delta_p$ and $\Delta_k$: for all $\Lambda_p \succeq 0$ and diagonal, $\Lambda_k \succeq 0$ and diagonal, $\B{v} \in \mathcal{L}_{2e}$, $\B{w} = \Delta(\B{v})$, and $T \geq 0$,
\begin{align}
    \int_0^T \left. \begin{bmatrix} \B{v}(t) \\ \B{w}(t) \end{bmatrix}^\top M \begin{bmatrix} \B{v}(t) \\ \B{w}(t) \end{bmatrix} \right. \geq 0
\end{align}
where
\begin{align*}
    M & \triangleq 
    \begin{bmatrix} M_{v v} & M_{v w} \\ M_{v w}^\top & M_{w w} \end{bmatrix} 
    = \left[\begin{array}{cc|cc}
        \alpha^2 \Lambda_p & 0 &0 & 0 \\
        0 & 0 & 0 & \Lambda_k \\ \hline
        0 & 0 & -\Lambda_p & 0 \\
        0 & \Lambda_k & 0 & -2\Lambda_k
    \end{array}\right].
\end{align*}

We now present the stability condition for the closed-loop system \eqref{eq:closed-loop}.

\begin{lem} \label{lemma:certification}
    Let $\alpha \geq 0$, $\sigma \in \mathbb{R}$, the plant in \eqref{eq:plant}, and the controller in \eqref{eq:controller} be given.
    If there exist $X \in \mathbb{R}^{(n_p + n_k) \times (n_p + n_k)}$ with $X \succ 0$, $\Lambda_p \in \mathbb{R}^{n_u \times n_u}$ with $\Lambda_p \succeq 0$ and diagonal, and $\Lambda_k \in \mathbb{R}^{n_\phi \times n_\phi}$ with $\Lambda_k \succeq 0$ and diagonal, such that
    \begin{align} \label{eq:certification-lmi}
        \begin{bmatrix}
            A^\top X + X A & X B_w \\ B_w^\top X & 0
        \end{bmatrix}
        +
        (\star)^\top M \begin{bmatrix}
            C_v & D_{v w} \\
            0 & I
        \end{bmatrix}
        \preceq 0,
    \end{align}
    then the controller satisfies the disk margin $D(\alpha, \sigma)$.
\end{lem}
\proof{
    Since the closed-loop system is in the form of an LTI system interconnected with an uncertainty described by an integral quadratic constraint, we apply Lemma 1 of \cite{junnarkar2024synthesizingneuralnetworkcontrollers}.$\hfill\blacksquare$
}
\begin{rem}
    This condition can also be used to find the largest $\alpha$ the closed-loop system satisfies by bisection on $\alpha$.
\end{rem}

\section{Neural Network Training} \label{sec:training}

\begin{table*}[bthp]
\begin{equation} \label{eq:expanded-lmi}
    \left[
    \begin{array}{cc|cc|c}
        \mathrm{He}(A_p R + B_p N_{A21}) & A_p + B_p N_{A22} C_p + N_{A11}^\top
        & B_p & B_p D_{kuw} + N_C^\top
        & \alpha N_{A21}^\top \tilde{L}_{\Delta_p}^\top
        \\
        N_{A11} + A_p^\top + C_p^\top N_{A22}^\top B_p^\top &  \mathrm{He}(S A_p + N_{A12} C_p)
        & S B_p & N_B + C_p^\top \hat{D}_{kvy}
        & \alpha C_p^\top N_{A22}^\top \tilde{L}_{\Delta_p}^\top
        \\
        \hline
        B_p^\top & B_p^\top S
        & -\Lambda_p & 0
        & \frac{\alpha(1+\sigma)}{2} \tilde{L}_{\Delta_p}^\top
        \\
        D_{kuw}^\top B_p^\top + N_C & N_B^\top + \hat{D}_{kvy} C_p
        & 0 & \hat{D}_{kvw} + \hat{D}_{kvw}^\top - 2\Lambda_k
        & \alpha D_{kuw}^\top \tilde{L}_{\Delta_p}^\top
        \\
        \hline
        \alpha \tilde{L}_{\Delta_p} N_{A21} & \alpha \tilde{L}_{\Delta_p} N_{A22} C_p
        & \frac{\alpha(1+\sigma)}{2} \tilde{L}_{\Delta_p} & \alpha \tilde{L}_{\Delta_p} D_{kuw}
        & -I
    \end{array}
    \right]
    \preceq 0
\end{equation}
\end{table*}

To synthesize a neural network controller that satisfies a specified disk margin $D(\alpha, \sigma)$, we use a reinforcement learning procedure that alternates between a training step and a stability margin-enforcing step.
The training step improves the controller's performance on the reward function, but the new controller does not necessarily satisfy the desired disk margin.
Therefore, we follow the training step with a stability margin-enforcing step that projects the controller into the set of controllers guaranteed to robustly stabilize the plant.

A direct application of Lemma~\ref{lemma:certification} gives the following projection problem, with $\theta^\prime$ being the controller parameters we would like to project into the set of controller parameters that satisfy the robust stability condition:
\begin{align*}
    \min_{\theta, X, \Lambda_p, \Lambda_k} &\quad \| \theta^\prime - \theta\| \\
    \text{s.t.} &\quad X \succ 0, \Lambda_p \succeq 0, \Lambda_k \succeq 0 \\
    &\quad \Lambda_p \ \text{and}\ \Lambda_k\ \text{diagonal} \\
    &\quad \eqref{eq:certification-lmi}\ \text{holds with}\ \theta.
\end{align*}
However, this is not computationally tractable since it is not convex in $\theta, X, \Lambda_p,$ and $\Lambda_k$.
This is due to bilinear terms in $\theta$ and $X$ that appear in \eqref{eq:certification-lmi} (for example, in $A^\top X$) and due to quadratic terms in $\theta$ and $\Lambda_p$ and in $\theta$ and $\Lambda_k$.
Therefore, if using Lemma~\ref{lemma:certification} directly, we would project $\theta^\prime$ into the following set parameterized by $X$, $\Lambda_p$, and $\Lambda_k$:
\begin{align} \label{eq:Theta-set}
    \Theta(X, \Lambda_p, \Lambda_k) \triangleq \left\{
        \theta | \eqref{eq:certification-lmi}\ \text{holds}
    \right\}.
\end{align}

To reduce conservatism, we leverage the transformation procedure from \cite{junnarkar2024synthesizingneuralnetworkcontrollers}.
Below we provide the resulting matrix inequality that is equivalent to Lemma~\ref{lemma:certification} when $\Lambda_k \succ 0$ but jointly convex in $\theta$, $X$, and $\Lambda_k$---leaving only $\Lambda_p$ fixed during the projection step.
For simplicity, we assume the order of the controller is the same as the order of the plant.

\begin{lem} \label{lemma:synthesis}
    Let the disk margin $D(\alpha, \sigma)$, the plant in \eqref{eq:plant}, and $\Lambda_p = \tilde{L}_{\Delta_p}^\top \tilde{L}_{\Delta_p}$, where $\Lambda_p \succeq 0$ and diagonal, be given.
    If there exist $\hat{\theta} = \left\{S, R, N_A, N_B, N_C, D_{k u w}, \hat{D}_{k v y}, \hat{D}_{k v w}, \Lambda_k\right\}$ where $S \succ 0$, $R \succ 0$, $\Lambda_k \succ 0$ and diagonal, and $\begin{bsmallmatrix}
        R & I \\ I & S
    \end{bsmallmatrix} \succ 0$ such that \eqref{eq:expanded-lmi} holds, then a controller $\theta$ can be reconstructed from $\hat{\theta}$ such that the controller satisfies the specified disk margin.
\end{lem}

We label the set of $\hat{\theta}$ that satisfy the conditions of Lemma~\ref{lemma:synthesis} for a particular $\Lambda_p$ as $\hat{\Theta}(\Lambda_p)$.
Given $\hat{\theta} \in \hat{\Theta}(\Lambda_p)$, the controller $\theta$ can be reconstructed by solving the following for $A_k, B_{kw}, B_{ky}, C_{kv}, D_{kvw}, D_{kvy}, C_{ku}, D_{kuy}$, where $U$ and $V$ are such that $VU^\top = I - RS$ (for example, by taking the SVD):
\begin{align*}
    N_A &= \begin{bmatrix} S A_p R & 0 \\ 0 & 0 \end{bmatrix} \\& \hphantom{=} + \begin{bmatrix}
        U & S B_p \\ 0 & I
    \end{bmatrix} \begin{bmatrix}
        A_k & B_{ky} \\ C_{ku} & D_{kuy}
    \end{bmatrix} \begin{bmatrix}
        V^\top & 0 \\ C_{py} R & I
    \end{bmatrix} \\
    N_B & = S B_p D_{kuw} + U B_{kw}\\
    N_C & = \Lambda_k D_{kvy} C_p R + \Lambda_k C_{kv}V^\top \\
    \hat{D}_{kvy} & = \Lambda_k D_{kvy}, \quad 
    \hat{D}_{kvw}  = \Lambda_k D_{kvw}.
\end{align*}

This controller is certified by the given $\Lambda_p$, the $\Lambda_k$ decision variable, and the following construction for $X$:
\begin{align*}
    X \triangleq \begin{bmatrix}
        I & S \\ 0 & U^\top
    \end{bmatrix} \begin{bmatrix}
        R & I \\ V^\top & 0
    \end{bmatrix}^{-1}.
\end{align*}
Therefore, satisfaction of Lemma~\ref{lemma:synthesis} implies $\Theta(X, \Lambda_p, \Lambda_k)$ is nonempty for the above $X$, $\Lambda_p$, and $\Lambda_k$.

Note that the transformation between $\hat{\theta}$ and $\theta$ also works in the reverse: a $\hat{\theta}$ can be constructed from a $\theta$, $X$, $\Lambda_p$, and $\Lambda_k$ using the above equations.

Critically, all matrix inequalities in Lemma~\ref{lemma:synthesis} are affine in $\hat{\theta}$, so these conditions can be incorporated into the following semidefinite program for projection:
\begin{equation} \label{eq:projection}
\begin{aligned}
    \min_{\hat{\theta}} &\quad \| \hat{\theta}^\prime - \hat{\theta}\| \\
    \text{s.t.} &\quad S \succ 0, R \succ 0, \Lambda_k \succ 0, \begin{bsmallmatrix}
        R & I \\ I & S
    \end{bsmallmatrix} \succ 0 \\
    &\quad \Lambda_k\ \text{diagonal} \\
    &\quad \eqref{eq:expanded-lmi} \ \text{holds}.
\end{aligned}
\end{equation}

With the above methods to certify a disk margin, to construct $\hat{\theta}$, and to project a $\hat{\theta}$ into $\hat{\Theta}(\Lambda_p)$, we introduce the following training algorithm.

\begin{algorithm}
    \caption{Neural Network Controller Training}
    \begin{algorithmic}[1]
        \State $\theta \gets$ arbitrary
        \State $X, \Lambda_p, \Lambda_k \gets I$ \label{algline:init}
        \For{$i=1,\dots$}
            \LineComment{Reinforcement learning step}
            \State $\theta^\prime \gets \Call{ReinforcementLearningStep}{\theta}$
            \LineComment{Robust stability-enforcing step}
            \If{$\exists X^\prime, \Lambda_p^\prime, \Lambda_k^\prime : \theta^\prime$ satisfies disk margins} \label{algline:dissipativity-check}
                \State $\theta, X, \Lambda_p, \Lambda_k \gets \theta^\prime, X^\prime, \Lambda_p^\prime, \Lambda_k^\prime$
            \Else
                \State $\hat{\theta}^\prime \gets \Call{ConstructThetaHat}{\theta^\prime, X, \Lambda_k}$ \label{algline:projection-start}
                \State $\hat{\theta} \gets \Call{ThetaHatProject}{\hat{\theta}^\prime, \hat{\Theta}(\Lambda_p)}$
                \State $X, \Lambda_k \gets \Call{ExtractFrom}{\hat{\theta}}$
                \State $\theta \gets \arg\min_{\theta} \|\theta - \theta^\prime\| : \theta \in \Theta(X, \Lambda_p, \Lambda_k)$ \label{algline:theta-proj}
            \EndIf
        \EndFor
    \end{algorithmic}
    \label{alg:training}
\end{algorithm}

Algorithm~\ref{alg:training} alternates between improving the controller through reinforcement learning, and modifying the controller to ensure that the controller satisfies the specified disk margins.
While a direct application of the projection in \eqref{eq:projection} fixes $\Lambda_p$, we introduce alternation on $\Lambda_p$ during training by, in line \ref{algline:dissipativity-check}, certifying the disk margin if possible.
This both allows $\Lambda_p$ to vary, and avoids the more expensive projection procedure when possible.
If the controller after the training step cannot be certified to satisfy the disk margin, we use the most recent $X$ and $\Lambda_k$ to transform it into the $\hat{\theta}$-space.
Then, we leverage the projection in \eqref{eq:projection} to construct $X$ and $\Lambda_k$ such that $\Theta(X, \Lambda_p, \Lambda_k)$ is nonempty.
Finally, we project $\theta^\prime$ into this set.
We use this projection in $\theta$-space instead of directly transforming the projected $\hat{\theta}$ into a $\theta$ since our goal is to find the closest $\theta \in \Theta(X, \Lambda_p, \Lambda_k)$ to $\theta^\prime$, and the procedure to recover $\theta$ from $\hat{\theta}$ gives only \textit{some} $\theta \in \Theta(X, \Lambda_p, \Lambda_k)$.

\begin{rem}
    One useful choice for initializing parameters $\theta$, $X$, and $\Lambda_p$ in Algorithm~\ref{alg:training} is an LTI controller synthesized to stabilize, or robustly stabilize, the plant.
\end{rem}

\section{Numerical Experiments} \label{sec:experiments}

\begin{figure}[tbp]
    \centering
    \includegraphics[width=0.4\linewidth]{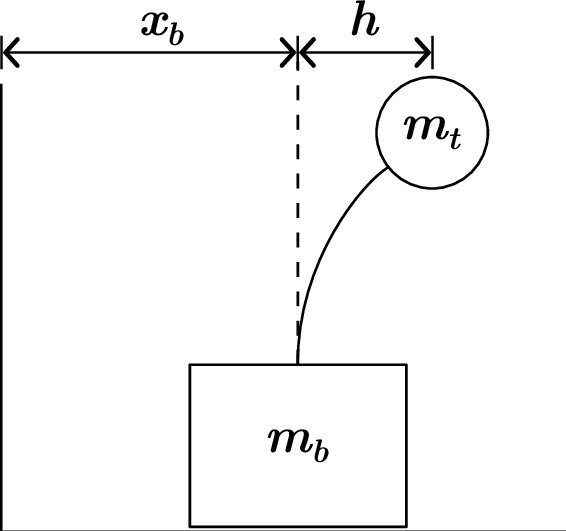}
    \caption{Diagram of the flexible rod on a cart.}
    \label{fig:flexarm-diagram}
\end{figure}

We demonstrate the performance of our method through simulation on a flexible rod on a cart.
We compare our method with two unconstrained neural network controllers and an LTI controller.
For shorthand, we refer to our method as Stability Margin Recurrent Implicit Neural Network (SM-RINN).
The unconstrained neural network controllers we compare against are an unconstrained RINN (U-RINN) and a standard fully connected feedforward neural network (FCNN).
The LTI controller (referred to as LTI) is trained using the same method as the SM-RINN: it guarantees the same stability margins.

Since all plants and controller models are defined in continuous time, we implement them by simulating between timesteps with the Runge-Kutta 4th order method.
We use PyTorch and Ray RLLib with the proximal policy optimization algorithm for training.
All code is online at { \href{https://github.com/neelayjunnarkar/nn-stability-margins}{github.com/neelayjunnarkar/nn-stability-margins}}. 

The full model of the flexible rod on a cart is as follows:
\begin{equation} \label{eq:flexarm-flexible-model}
    \begin{aligned}
        \dot{\B{x}}_{\B{f}}(t) & = \begin{bmatrix}
            0 & I \\ -M^{-1} K & -M^{-1} B
        \end{bmatrix} \B{x_f}(t)
        + \begin{bmatrix}
            0 \\ M^{-1} \begin{bsmallmatrix}
                1 \\ 0
            \end{bsmallmatrix}
        \end{bmatrix} \B{u}(t) \\
        \B{y}(t) & = \begin{bmatrix}
            1 & 1 & 0 & 0 
        \end{bmatrix} \B{x_f}(t) \\
        M & = \begin{bmatrix}
            m_b + m_r + \rho L & m_t + \frac{\rho L}{3}\\ m_t + \frac{\rho L}{3} & m_t + \frac{\rho L}{5}
        \end{bmatrix} \\
        K & = \begin{bmatrix}
            0 & 0 \\ 0 & \frac{4EI}{L^3}
        \end{bmatrix}, \quad
        B = \begin{bmatrix}
            0 & 0 \\ 0 & 0.9
        \end{bmatrix}
    \end{aligned}
\end{equation}
where $x_f(t) = \begin{bmatrix} x_b(t) & h(t) & \dot{x_b}(t) & \dot{h}(t) \end{bmatrix}^\top \in \mathbb{R}^4$ is the state of the flexible rod on a cart, $x_b(t)$ is the position of the rod's base, $h(t)$ is the horizontal deviation of the top of the rod from the base, $m_b = 1$ kg is the mass of the base, $m_t = 0.1$ kg is the mass of an object at the top of the rod, $L = 1$ m is the rod length, $\rho = 0.1$ N/m is the mass density of the rod, $r = 10^{-2}$ m is the radius of the rod cross-section, $E = 200 \cdot 10^{-9}$ GPa is the Young's modulus of steel, and $I = \frac{\pi}{4} r^4$ m$^4$ is the area second moment of inertia.
The output $y$ represents the measurement of the position of the rod tip.

The flexible model is used in simulation for training.
As a demonstration of using a simpler model for design, we use the following rigid model of the rod on a cart for all stability margin properties of the SM-RINN and LTI controllers:
\begin{equation} \label{eq:flexarm-rigid-model}
     \begin{aligned}
        \dot{\B{x}}_{\B{r}}(t) & = \begin{bmatrix}
            0 & 1 \\ 0 & 0
        \end{bmatrix} \B{x_r}(t) 
        + \begin{bmatrix}
            0 \\ \frac{1}{m_b + m_r + \rho L}
        \end{bmatrix} \B{u}(t)
        \\
        \B{y}(t) & = \B{x_b}(t) \\
    \end{aligned}
\end{equation}
where $x_r(t) = \begin{bmatrix} x_b(t) & \dot{x}_b(t)\end{bmatrix}^\top$.
For the SM-RINN and LTI controllers, we impose a requirement that they satisfy a disk margin with $\alpha=0.353$ and $\sigma=0$.
The value of $\alpha$ can be tuned to tradeoff between reward and robustness.
This particular disk margin implies a lower bound on the gain margin of 3dB and a lower bound on the phase margin of $20^\circ$.
Note that disk margins ensure robustness to simultaneous variation in gain and phase as well, which gain and phase margins do not consider.

We simulate the controllers and plant with a time step of $0.001$ seconds, and train with a time horizon of 2 seconds.
The reward for each time step is $\exp(-\|x_f(t)\|^2) + \exp(-u(t)^2)$.
Initial conditions are sampled uniformly at random with $x_b(0) \in [-1, 1]$ m, $h(0) \in [-0.44, 0.44]$ m, $\dot{x}_b(0) \in [-0.25, 0.25]$ m/s, and $\dot{h}(0) \in [-2, 2]$ m/s.
The control input is saturated to the interval $[-20, 20]$ N.

For the two RINN controllers, we use a state size of 2 ($n_k = 2$) and 16 activation functions ($n_\phi = 16$).
For the FCNN, we use two layers of size 19 each to match the total number parameters of the RINN controllers.
The U-RINN, FCNN, and LTI controllers are trained with a learning rate of $10^{-4}$.
The SM-RINN is trained with a learning rate of $5 \cdot 10^{-5}$ to alleviate instability.
The activation function is $\tanh$ for all neural network controllers.

\begin{figure}[t]
    \centering
    \includegraphics[width=0.9\linewidth]{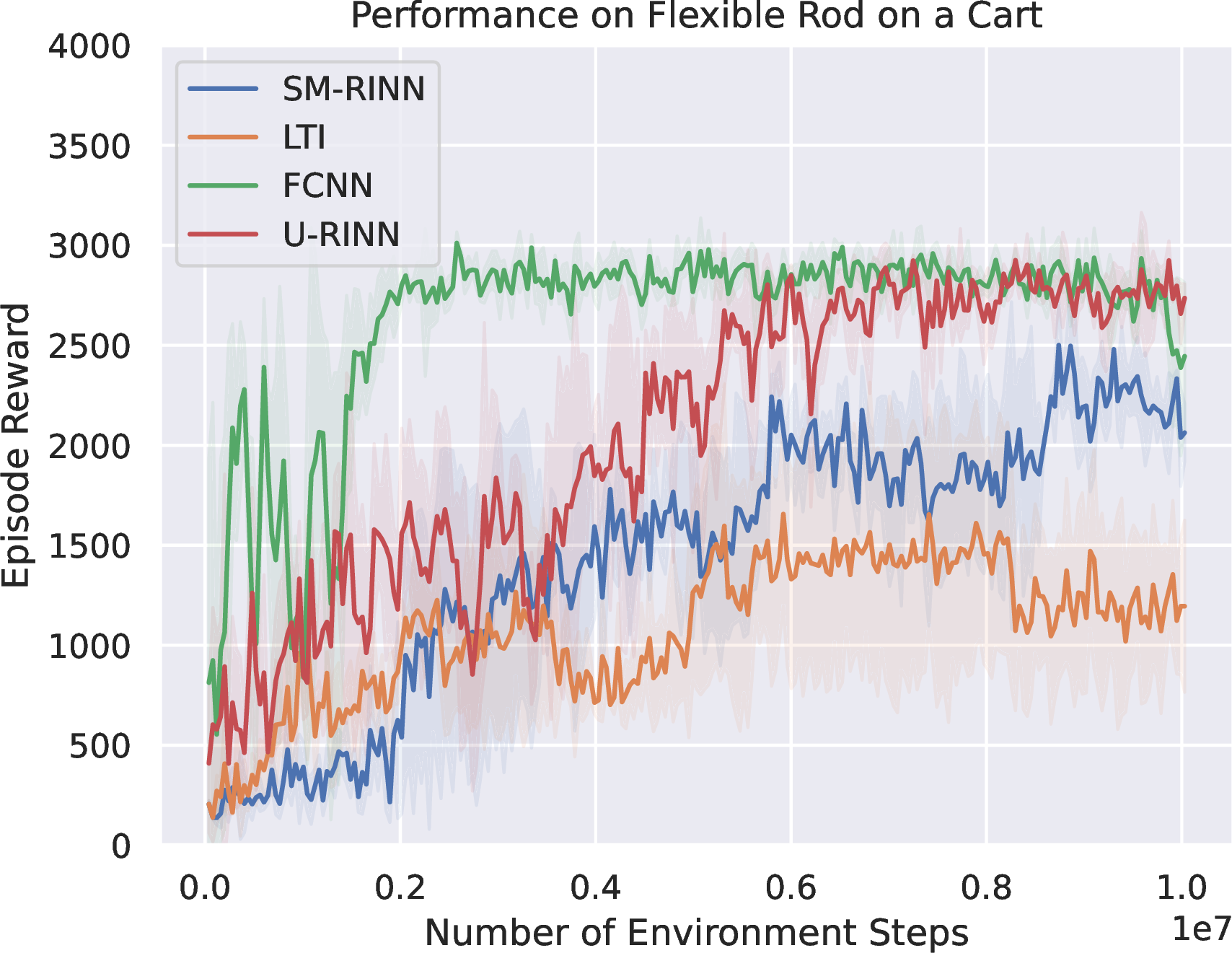}
    \caption{
    Evaluation reward vs. number of training environment steps for the Stability Margin Recurrent Implicit Neural Network (SM-RINN, our method), Fully Connected Neural Network (FCNN), Unconstrained RINN (U-RINN), and linear time-invariant (LTI) controllers on the flexible rod on a cart.
    The SM-RINN and LTI controllers both guarantee a disk margin of $\alpha=0.353$ with skew $\sigma = 0$.
    The solid line represents the mean, and the shaded region represents 1 standard deviation.
    The reward is upper bounded by 4000, and lower bounded by 0.
    Higher reward indicates a combination of lower control effort and smaller state of the flexible rod on a cart.
    }
    \label{fig:flexarm}
\end{figure}

Figure~\ref{fig:flexarm} shows reward vs. number of training environment steps sampled, with mean and standard deviations taken over three runs of each controller.
The two controllers with the highest rewards are the unconstrained controllers (FCNN and U-RINN, both about 3000).
The next highest is  the SM-RINN (reward about 2250), which guarantees stability margins.
The LTI controller, which guarantees the same stability margins as the SM-RINN, achieves the lowest reward (about 1500).
This ordering shows a typical performance vs. robustness tradeoff: the methods without robustness requirements achieve the highest reward.
Note that our method, the SM-RINN, achieves a significant reward improvement over the LTI controller while guaranteeing the same margin.



\section{Conclusion} \label{sec:conclusion}

We presented a method to analyze the disk margin of a neural network controller, and a method to train a neural network controller to maximize reward subject to it satisfying a specified disk margin.
We demonstrated our approach on a flexible rod on a cart, where it achieved significantly higher reward than did an LTI controller.
Our method enables neural networks to be used as controllers in safety-critical applications where stability margins are required.









\renewcommand*{\bibfont}{\normalfont\small}
\printbibliography

@inproceedings{junnarkarSynthesisStabilizingRecurrent2022,
  title = {Synthesis of {{Stabilizing Recurrent Equilibrium Network Controllers}}},
  booktitle = {2022 {{IEEE}} 61st {{Conference}} on {{Decision}} and {{Control}} ({{CDC}})},
  author = {Junnarkar, Neelay and Yin, He and Gu, Fangda and Arcak, Murat and Seiler, Peter},
  date = {2022-12},
  eventtitle = {2022 {{IEEE}} 61st {{Conference}} on {{Decision}} and {{Control}} ({{CDC}})}
}

@article{seilerIntroductionDiskMargins2020,
  title = {An {{Introduction}} to {{Disk Margins}} [{{Lecture Notes}}]},
  author = {Seiler, Peter and Packard, Andrew and Gahinet, Pascal},
  date = {2020-10},
  journaltitle = {IEEE Control Systems Magazine},
  volume = {40},
  number = {5},
  eventtitle = {{{IEEE Control Systems Magazine}}}
}

@article{schugRobustnessMarginsLinear2017,
  title = {Robustness {{Margins}} for {{Linear Parameter Varying Systems}}},
  author = {Schug, Ann-Kathrin and Seiler, Peter and Pfifer, Harald},
  date = {2017},
  journaltitle = {AerospaceLab Journal},
  volume = {Issue 13},
  publisher = {ONERA},
  langid = {english}
}

@book{zhouRobustOptimalControl1996,
  title = {Robust and {O}ptimal {C}ontrol},
  author = {Zhou, Kemin and Doyle, John C. and Glover, Keith},
  date = {1996},
  publisher = {Prentice-Hall, Inc.},
  location = {USA},
}

@article{elghaouiImplicitDeepLearning2021,
  title = {Implicit {{Deep Learning}}},
  author = {El Ghaoui, Laurent and Gu, Fangda and Travacca, Bertrand and Askari, Armin and Tsai, Alicia},
  date = {2021-01},
  journaltitle = {SIAM Journal on Mathematics of Data Science},
  volume = {3},
  number = {3},
  publisher = {{Society for Industrial and Applied Mathematics}},
}

@article{Gu_Yin_Ghaoui_Arcak_Seiler_Jin_2022,
  title = {Recurrent {N}eural {N}etwork {C}ontrollers {S}ynthesis with {S}tability {G}uarantees for {P}artially {O}bserved {S}ystems},
  author = {Gu, Fangda and Yin, He and Ghaoui, Laurent El and Arcak, Murat and Seiler, Peter and Jin, Ming},
  date = {2022-06},
  journaltitle = {Proceedings of the AAAI Conference on Artificial Intelligence},
}

@book{callierLinearSystemTheory1991,
  title = {Linear {{System Theory}}},
  author = {Callier, Frank M. and Desoer, Charles A.},
  editor = {Thomas, John B.},
  editortype = {redactor},
  date = {1991},
  series = {Springer {{Texts}} in {{Electrical Engineering}}},
  publisher = {Springer},
  location = {New York, NY},
}

@article{wangLearningAllStabilizing2023,
  title = {Learning {{Over All Stabilizing Nonlinear Controllers}} for a {{Partially-Observed Linear System}}},
  author = {Wang, Ruigang and Barbara, Nicholas H. and Revay, Max and Manchester, Ian R.},
  date = {2023},
  journaltitle = {IEEE Control Systems Letters},
  volume = {7},
  eventtitle = {{{IEEE Control Systems Letters}}}
}

@misc{junnarkar2024synthesizingneuralnetworkcontrollers,
      title={Synthesizing {N}eural {N}etwork {C}ontrollers with {C}losed-{L}oop {D}issipativity {G}uarantees}, 
      author={Neelay Junnarkar and Murat Arcak and Peter Seiler},
      year={2024},
      eprint={2404.07373},
      archivePrefix={arXiv},
      primaryClass={eess.SY},
}

@article{megretskiSystemAnalysisIntegral1997,
  title = {System {{Analysis Via Integral Quadratic Constraints}}},
  author = {Megretski, A. and Rantzer, A.},
  date = {1997-06},
  journaltitle = {IEEE Transactions on Automatic Control},
  volume = {42},
  number = {6},
  eventtitle = {{{IEEE Transactions}} on {{Automatic Control}}}
}

@article{seilerStabilityAnalysisDissipation2015,
  title = {Stability {{Analysis With Dissipation Inequalities}} and {{Integral Quadratic Constraints}}},
  author = {Seiler, Peter},
  date = {2015-06},
  journaltitle = {IEEE Transactions on Automatic Control},
  volume = {60},
  number = {6},
  eventtitle = {{{IEEE Transactions}} on {{Automatic Control}}}
}

@article{veenmanIQCsynthesisGeneralDynamic2014,
  title = {{{IQC-synthesis}} with {G}eneral {D}ynamic {M}ultipliers},
  shorttitle = {{{IQC-synthesis}} with General Dynamic Multipliers},
  author = {Veenman, Joost and Scherer, Carsten W.},
  date = {2014-11-25},
  journaltitle = {International Journal of Robust and Nonlinear Control},
  shortjournal = {Int. J. Robust. Nonlinear Control},
  volume = {24},
  number = {17},
  langid = {english}
}

@article{schererDissipativityIntegralQuadratic2022,
  title = {Dissipativity and {{Integral Quadratic Constraints}}: {{Tailored Computational Robustness Tests}} for {{Complex Interconnections}}},
  shorttitle = {Dissipativity and {{Integral Quadratic Constraints}}},
  author = {Scherer, Carsten W.},
  date = {2022-06},
  journaltitle = {IEEE Control Systems Magazine},
  volume = {42},
  number = {3},
  pages = {115--139},
  eventtitle = {{{IEEE Control Systems Magazine}}}
}

@article{fazlyabSafetyVerificationRobustness2022,
  title = {Safety {{Verification}} and {{Robustness Analysis}} of {{Neural Networks}} via {{Quadratic Constraints}} and {{Semidefinite Programming}}},
  author = {Fazlyab, Mahyar and Morari, Manfred and Pappas, George J.},
  date = {2022-01},
  journaltitle = {IEEE Transactions on Automatic Control},
  volume = {67},
  number = {1},
  eventtitle = {{{IEEE Transactions}} on {{Automatic Control}}}
}

@article{yinStabilityAnalysisUsing2022,
  title = {Stability {{Analysis Using Quadratic Constraints}} for {{Systems With Neural Network Controllers}}},
  author = {Yin, He and Seiler, Peter and Arcak, Murat},
  date = {2022-04},
  journaltitle = {IEEE Transactions on Automatic Control},
  volume = {67},
  number = {4},
  eventtitle = {{{IEEE Transactions}} on {{Automatic Control}}}
}

@inproceedings{hashemiCertifyingIncrementalQuadratic2021,
  title = {Certifying {{Incremental Quadratic Constraints}} for {{Neural Networks}} via {{Convex Optimization}}},
  booktitle = {Proceedings of the 3rd {{Conference}} on {{Learning}} for {{Dynamics}} and {{Control}}},
  author = {Hashemi, Navid and Ruths, Justin and Fazlyab, Mahyar},
  date = {2021-05-29},
  publisher = {PMLR},
  eventtitle = {Learning for {{Dynamics}} and {{Control}}},
  langid = {english}
}

@inproceedings{pauliLinearSystemsNeural2021,
  title = {Linear {S}ystems with {N}eural {N}etwork {N}onlinearities: {{Improved}} {S}tability {A}nalysis via {A}causal {{Zames-Falb}} {M}ultipliers},
  shorttitle = {Linear Systems with Neural Network Nonlinearities},
  booktitle = {2021 60th {{IEEE Conf.}} on {{Decision}} and {{Control}} ({{CDC}})},
  author = {Pauli, Patricia and Gramlich, Dennis and Berberich, Julian and Allgöwer, Frank},
  date = {2021-12},
  eventtitle = {2021 60th {{IEEE Conference}} on {{Decision}} and {{Control}} ({{CDC}})}
}

@incollection{karnyRecurrentNeuralNetworks1998,
  title = {Recurrent {{Neural Networks}}: {{Some Systems-Theoretic Aspects}}},
  shorttitle = {Recurrent {{Neural Networks}}},
  booktitle = {Dealing with {{Complexity}}: {{A Neural Networks Approach}}},
  author = {Sontag, Eduardo},
  date = {1998},
  publisher = {Springer},
  location = {London},
  langid = {english}
}

@misc{faa,
    title={Code of {F}ederal {R}egulations. {T}itle 14: {C}hapter 1: Subchapter {C}: Part 25: Airworthiness Standards: Transport Category Airplanes},
    year={2023},
}

@article{barabanovStabilityAnalysisDiscretetime2002,
  title = {Stability Analysis of Discrete-Time Recurrent Neural Networks},
  author = {Barabanov, N.E. and Prokhorov, D.V.},
  date = {2002-03},
  journaltitle = {IEEE Transactions on Neural Networks},
  volume = {13},
  number = {2},
  pages = {292--303},
  eventtitle = {{{IEEE Transactions}} on {{Neural Networks}}}
}

\end{document}